\documentclass[11pt]{article}
\usepackage[usenames,dvipsnames]{pstricks}
\usepackage{times}
\usepackage{graphicx}
\usepackage{amsmath}
\usepackage{amssymb}
\usepackage{amsthm}
\usepackage{amsfonts}
\usepackage{latexsym}
\usepackage{mathrsfs}
\usepackage[mathcal]{euscript}

\newtheorem{lemma}{Lemma}

\newtheorem{theorem}{Theorem}

\def\G{\mathrm G}

\def\X{\mathrm X}

\begin{document}
\title{A 4/3-approximation for TSP on cubic 3-edge-connected graphs} 
\author{Nishita Aggarwal  \and   Naveen Garg    \and      Swati Gupta}
\maketitle
\section{Introduction}
We consider the travelling salesman problem on metrics which can be viewed as the shortest path metric of an undirected graph with unit edge-lengths. Finding  a TSP tour in such a metric is then equivalent to finding a connected Eulerian subgraph in the underlying graph. Since the length of the tour is the number of edges in this Eulerian subgraph our problem can equivalently be stated as follows: Given an undirected, unweighted graph $G=(V,E)$ find a connected Eulerian subgraph, $H=(V,E')$ with the fewest edges. Note that $H$ could be a multigraph. 

In this paper we consider the special case of the problem when $G$ is 3-regular (also called cubic) and 3-edge-connected. Note that the smallest Eulerian subgraph contains at least $n=|V|$ edges. In fact, in the shortest path metric arising out of such a graph the Held-Karp bound for the length of the TSP tour would also be $n$. This is because we can obtain a fractional solution to the sub-tour elimination LP (which is equivalent to the Held-Karp bound) of value $n$ by assigning 2/3 to every edge in $G$. 

Improving the approximation ratio for metric-TSP beyond 3/2 is a long standing open problem. For the metric completion of cubic 3-edge connected graphs Gamarnik et.al.~\cite{gamarnik} obtained an algorithm with an approximation guarantee slightly better than 3/2. The main result of this paper is to improve this approximation guarantee to 4/3 by giving a polynomial time algorithm to find a connected Eulerian subgraph with at most 4n/3 edges. This matches the conjectured integrality gap for the sub-tour elimination LP for the special case of these metrics.
 
\section{Preliminaries}
Let $n$ be the number of vertices of the given graph G. Let $d(x)$ denote the degree of x. A 2-factor in $G$ is a subset of edges $X$ such that every vertex has degree 2 in $X$. Let $\sigma(\X)$ denote the minimum size of components of X. Given two distinct edges $e_1=x_1v$ and $e_2=x_2v$ incident on a vertex $v$, let ${G_v}^{e_1,e_2}$ denote the graph obtained by replacing $e_1,e_2$ by the edge $x_1x_2$. The vertex $v$ is said to be {\it split off}. We call a cut $(S,\overline{S})$ {\it essential} when both $S$ and $\overline{S}$ contain at least one edge each. 

We will need the following results for our discussion
\begin{lemma}[Peterson\cite{Petersen}]
Every bridgeless cubic graph has a 2-factor.
\end{lemma}
\begin{lemma}[Mader\cite{mader}]
\label{mader}
Let $G=(V,E)$ be a $k$-edge-connected graph, $v\in V$ with $d(v) \geq k+2$. Then there exists edges $e_1,e_2\in E$ such that $G_v^{e_1,e_2}$ is homeomorphic to a $k$-edge-connected graph.
\end{lemma}
\begin{lemma}[Jackson, Yoshimoto\cite{JY06}]
\label{JY}
Let $G$ be a 3-edge-connected graph with $n$ vertices.
Then $G$ has a spanning even subgraph in which each component has at least $\min(n,5)$ vertices.
\end{lemma}
\section{Algorithm}
Our algorithm can be broadly split into three parts. We first find a 2-factor of the cubic graph that has no 3-cycles and 4-cycles. Next, we compress the 5-cycles into `super-vertices' and split them using Lemma 2 to get a cubic 3-edge-connected graph $\G^\prime$ again. Repeatedly applying the first part on $G^{\prime}$ and compressing the five cycles gives a 2-factor with no 5-cycle on the vertices of the original graph. We `expand' back the super-vertices to form $X$ that is a subgraph of G. We finally argue that $X$ can be modified to get a connected spanning even multi-graph using at most 4/3(n) edges. 
 
The starting point of our algorithm is Theorem~\ref{JY}~\cite{JY06}. In fact \cite{JY06} proves the following stronger theorem.
\begin{theorem}
\label{stronger}
Let $G$ be a 3-edge-connected graph with $n$ vertices, $u_2$ be a vertex of
$G$ with $d(u_2) = 3$, and $e_1 = u_1u_2; e_2 = u_2u_3$ be edges of $G$. (it may be the case that $u_1 = u_3$).
Then $G$ has a spanning even subgraph $X$ with $\left\{e_1,e_2\right\} \subset E(X)$ and $\sigma(X) \ge \min(n,5)$.
\end{theorem}
The proof of this theorem is non-constructive. We refer to the edges $e_1,e_2$ in the statement of the theorem as ``required edges". We now discuss the changes required in the proof given in \cite{JY06} to obtain a polynomial time algorithm which gives the subgraph $X$ with the properties as specified in Theorem~\ref{stronger}. Note that we will be working with a 3-regular graph (as against an arbitrary graph of min degree 3 in \cite{JY06}) and hence the even subgraph $X$ we obtain will be a 2-factor. 
\begin{enumerate}
\item If $G$ contains a non-essential 3-edge cut then we proceed as in the proof of Claim 2 in \cite{JY06}. This involves splitting $G$ into 2 graphs $G_1, G_2$ and suitably defining the required edges for these 2 instances so that the even subgraphs computed in these 2 graphs can be combined. This step is to be performed whenever the graph under consideration has an essential 3-edge cut. 
\item Since $G$ is 3-regular we do not require the argument of Claim 6.
\item Since $G$ has no essential 3-edge cut and is 3-regular, a 3-cycle in $G$ implies that $G$ is $K_4$. In this case we can find a spanning even subgraph containing any 2 required edges. 
\item The process of eliminating 4-cycles in the graph involves a sequence of graph transformations. The transformations are as specified in \cite{JY06} but the order in which the 4-cycles are considered depends on the number of required edges in the cycle. We first consider all such cycles which do not have any required edges, then cycles with 2 required edges and finally cycles which have one required edge. 

Since with each transformation the number of edges and vertices in the graph reduces we would eventually terminate with a graph, say $G'$, with girth 5. We find a 2-factor in $G'$, say $X'$ 
and undo the transformations (as specified in \cite{JY06}) in the reverse order in which they were done to obtain a 2-factor $X$ in the original graph $G$ which has the properties of Theorem~\ref{stronger}.
\end{enumerate}

Suppose the 2-factor obtained $X$ contains a 5-cycle $C$. We compress the vertices of $C$ into a single vertex, say $v_C$, and remove self loops. $v_C$ has degree 5 and we call this vertex a {\em super-vertex}. We now use Lemma~\ref{mader} to replace two edges $x_1v_C$ and $x_2v_C$ incident at $v_C$ with the edge $x_1x_2$ while preserving 3-edge connectivity. The edge $x_1x_2$ is called a {\em super-edge}. Since the graph obtained is cubic and 3-edge connected we can once again find a 2-factor, each of whose cycles has length at least 5. If there is a 5-cycle which does not contain any super-vertex or super-edge we compress it and repeat the above process. We continue doing this till we obtain a 2-factor, say $X$, each of whose cycles is either of length at least 6 or contains a super-vertex or a super-edge.

In the 2-factor $X$ we replace every super-edge with the corresponding edges. For instance the super-edge $x_1x_2$ would get replaced by edges $x_1v_C$ and $x_2v_C$ where $v_C$ is a super-vertex obtained by collapsing the vertices of a cycle $C$. After this process $X$ is no more a 2-factor but an even subgraph. However, the only vertices which have degree more than 2 are the super-vertices and they can have a maximum degree 4. Let $X$ denote this even subgraph.

Consider some connected component $W$ of $X$. We will show how to expand the super-vertices in $W$ into 5-cycles to form an Eulerian subgraph with at most $\lfloor 4|W'|/3\rfloor-2$ edges, where $|W'|$ is number of vertices in the expanded component. For each component we will use 2 more edges to connect this component to the other components to obtain a connected Eulerian subgraph with at most $\lfloor 4n/3\rfloor-2$ edges.  Note that the subgraph we obtain may use an edge of the original graph at most twice.

We now consider two cases depending on whether $W$ contains a super-vertex.
\begin{enumerate}
\item $W$ has no super-vertices. Then, $W$ is a cycle with at least 6 vertices and hence Eulerian. Since $|W|/3\ge 2$ the claim follows. 
\item $W$ has at least one super-vertex, say $s$. We will discuss the transformations for a single super-vertex and this will be repeated for the other super-vertices. Note that $s$ has degree 2 or 4. 
\begin{figure}[ht]
\begin{center}
\includegraphics[width=0.7\textwidth]{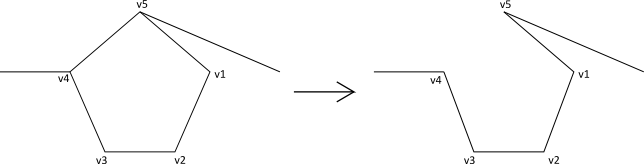}
\includegraphics[width=0.7\textwidth]{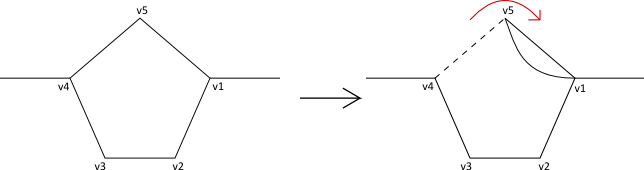}
\end{center}
\caption{\label{fig:degree2}On Expanding a super-vertex with degree 2}
\end{figure}

If $s$ has degree 2, then the 2 edges incident on the 5-cycle corresponding to $s$ would be as in Figure~\ref{fig:degree2}. In both cases we obtain an Eulerian subgraph. By this transformation we have added 4 vertices and at most 5 edges to the subgraph $W$. 

Suppose the super-vertex $s$ has degree 4 in the component $W$. $W$ may not necessarily be a component of the subgraph $X$ as it might have been obtained after expanding a few super-vertices, but that will not effect our argument. Let $C$ be the 5-cycle corresponding to this super-vertex and let $v_1,v_2,v_3,v_4,v_5$ be the vertices on $C$ (in order). Further let $v'_i$ be the vertex not in $C$ adjacent to $v_i$. Let $v_5v'_5$ be the edge incident on $C$ that is not in the subgraph $W$. 

\begin{figure}[ht]
\begin{center}
\includegraphics[width=0.6\textwidth]{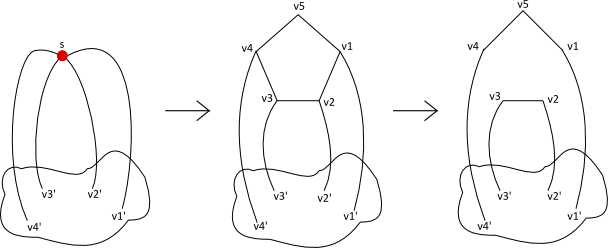}
\end{center}

\caption{\label{fig:conn4}Expanding a super-vertex with degree 4 when $v_1v_2$ and $v_3v_4$ do not form a 2-edge-cut of the sub-graph constructed till now.}
\end{figure}
\begin{figure}[ht]
\begin{center}
\includegraphics[width=0.5\textwidth]{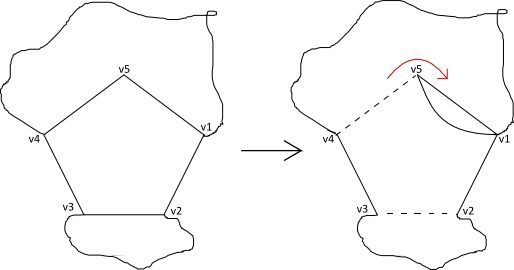}
\end{center}

\caption{\label{fig:v1v4}Expanding a super-vertex with degree 4 when $v_1v_2$ and $v_3v_4$ form a 2-edge-cut.}
\end{figure}
We replace the vertex $s$ in $W$ with the cycle $C$ and let $W'$ be the resulting subgraph. 
Note that by dropping edges $v_1v_2$ and $v_3v_4$ from $W'$ we obtain an Eulerian subgraph which includes all vertices of $C$. However, this subgraph may not be connected as it could be the case that edges $v_1v_2$ and $v_3v_4$ form an edge-cut in $W'$. If this is the case then we apply the transformation as shown in Figure~\ref{fig:v1v4}. This ensures that $W'$ remains connected and is Eulerian. Note that as a result of this step we have added 4 vertices and at most 4 edges to the subgraph $W$.
\end{enumerate}

Let $W'$ be the component obtained by expanding all the super-vertices in $W$. 
Suppose initially, component $W$ had $k_1$ super-vertices of degree 2, $k_2$ super-vertices of degree 4 and $k_3$ vertices of degree 2. This implies $W$ had $k_1$ + 2$k_2$ + $k_3$ edges. On expanding a super-vertex of degree 2, we add 5 edges in the worst case. On expanding a super-vertex of degree 4, we add 4 edges in the worst case. So, the total number of edges in $W'$ is at most  $6k_1+6k_2+k_3$ while the number of vertices in $W'$ is exactly $5k_1+5k_2+k_3$. Note that $k_1+k_2+k_3\ge 5$ and if $k_1+k_2+k_3=5$ then $k_1+k_2\ge 1$. Hence, $2k_1+2k_2+k_3\ge 6$ and this implies that the number of edges in $W'$ is at most $\lfloor 4|V(W')|/3\rfloor-2$.
 
\section{Conclusions}
We show that any cubic 3-edge connected graph contains a connected Eulerian subgraph with at most 4n/3 edges. It is tempting to conjecture the same for non-cubic graphs especially since the result in \cite{JY06} holds for all 3-edge connected graphs. The example of a $K_{3,n}$ demonstrates that this conjecture would be false. A $K_{3,n}$ is 3-edge connected and any connected Eulerian subgraph contains at least $2n$ edges.  
\bibliographystyle{plain}
\bibliography{main}

\end{document}